\documentclass[
 aps,prl,
 amsmath,amssymb,
reprint
]{revtex4-2}

\usepackage{graphicx}
\usepackage{dcolumn}
\usepackage{bm}
\usepackage{physics} 
\usepackage[utf8]{inputenc}
\usepackage[T1]{fontenc}
\usepackage{mathptmx}
\usepackage{etoolbox}
\usepackage{hyperref}
\usepackage{cleveref}
\usepackage{xcolor}
\usepackage{mathtools}
\usepackage{tikz}
\usepackage{float}
\usepackage{placeins}
\usepackage{amssymb}
\usepackage{amsmath}

\usetikzlibrary{arrows, positioning, math} 
\tikzset{
    >=stealth',
    punkt/.style={
           rectangle,
           rounded corners,
           draw=black, very thick,
           text width=6.5em,
           minimum height=2em,
           text centered},
    pil/.style={
           ->,
           thick,
           shorten <=2pt,
           shorten >=2pt,}
}

\makeatletter
\def\@email#1#2{
 \endgroup
 \patchcmd{\titleblock@produce}
  {\frontmatter@RRAPformat}
  {\frontmatter@RRAPformat{\produce@RRAP{*#1\href{mailto:#2}{#2}}}\frontmatter@RRAPformat}
  {}{}
}
\makeatother

\usepackage{mathtools}

\usepackage{nameref}

\begin{document}

\title{Memory Effects in Contact Line Friction}
\author{Niklas Wolf}
\email{wolf@cpc.tu-darmstadt.de} 
\affiliation{Department of Chemistry, Technical University of Darmstadt, 64287 Darmstadt, Germany}
\author{Nico F. A. van der Vegt}
\email{vandervegt@cpc.tu-darmstadt.de}
\affiliation{Department of Chemistry, Technical University of Darmstadt, 64287 Darmstadt, Germany}

\date{\today}

\begin{abstract}
When a drop of liquid comes into contact with a solid surface, it relaxes towards an equilibrium configuration, either wetting the surface or remaining in a droplet-like shape with a finite contact angle.
The force driving the process towards equilibrium is the corresponding out-of-balance Young's force.
However, the speed with which the liquid front advances depends strongly on an opposing friction force arising from dissipative processes due to the moving solid-liquid-gas contact line.
In analogy to the treatment of hydrodynamic friction we present an exact method, based on the Mori–Zwanzig formalism, to extract this friction from equilibrium data.
We find that the contact line exhibits long-lasting memory with a characteristic power-law decay due to coupling to the systems hydrodynamic modes.
Within linear response regime, we obtain the frequency-dependent dissipative and elastic response of the contact line to an external perturbation, including a frequency-dependent friction coefficient.
Similar to hydrodynamic friction in liquids, we find that the friction decreases beyond a characteristic frequency and the system exhibits predominantly elastic behavior.

\end{abstract}

\maketitle

\section{Introduction} \label{sec:intro}
Theories that describe dynamic wetting phenomena generally consider the microscopic motion of the contact line and the associated contact line friction to be Markovian.
An overlap between the relaxation of the liquid front and other microscopic relaxation processes is not considered.
For theories that employ continuum mechanics to describe dynamic wetting phenomena,\cite{hoffman_study_1975,voinov_hydrodynamics_1976, hocking_moving_1977, cox_dynamics_1986, pismen_mesoscopic_2002}, this is a natural choice.
However, molecular kinetic theory (MKT)\cite{blake_kinetics_1969,blake_influence_2002} (and related approaches,\cite{brochard-wyart_dynamics_1992,petrov_combined_1992}), which attribute the contact-line motion to molecular adsorption at the liquid front, similarly rely on a Markovian assumption.
From a comparison to hydrodynamic friction in fluids it could be argued that a Markovian description of the contact line starts to break down only at time scales comparable to the relaxation of molecular degrees of freedom.\cite{mason_optical_1995, gittes_microscopic_1997, squires_fluid_2010,straube_rapid_2020}

Herein, we demonstrate that the Markovian assumption breaks down at least one order of magnitude earlier, as the movement of the contact line and the contact line friction are dominated by persistent and long-lasting memory effects. 
We introduce a framework to obtain the full frequency-dependent contact-line friction, utilizing information from non-critical equilibrium fluctuations of the contact line, reducing to the approach by Toledano \textit{et al.}\cite{fernandez-toledano_contact-line_2019, fernandez-toledano_moving_2020} in the Markovian limit.
Within linear response theory, we then obtain the response of the contact line to an external force.
The dynamics of the contact line at short times are dominated by low-frequency molecular vibrations, with no modes that could be attributed to MKT absorption modes.
On long time scales, the contact-line correlation functions are governed by a power-law decay, due to coupling to the system’s hydrodynamic modes.
Notably, the majority of the contact-line friction emerges due to this coupling.

\section{Theoretical Background} \label{sec:theo}
At equilibrium, the equation of motion of the contact line may be written as a generalized Langevin equation (GLE)\cite{mori_transport_1965} in terms of the contact line position $X$ and velocity $V$
\begin{equation}
    m \frac{\dd V_t}{\dd t} = F_t = -kX_t-\int_0^{\infty}\dd s\, K(s)V_{t-s} + F^R_t, \label{eq:gle}
\end{equation}
where $K$ is the memory kernel, $F^R$ is a random force, $F$ is the force acting on the contact line, $k$ is a harmonic force constant and $m$ is the mass of the contact line.
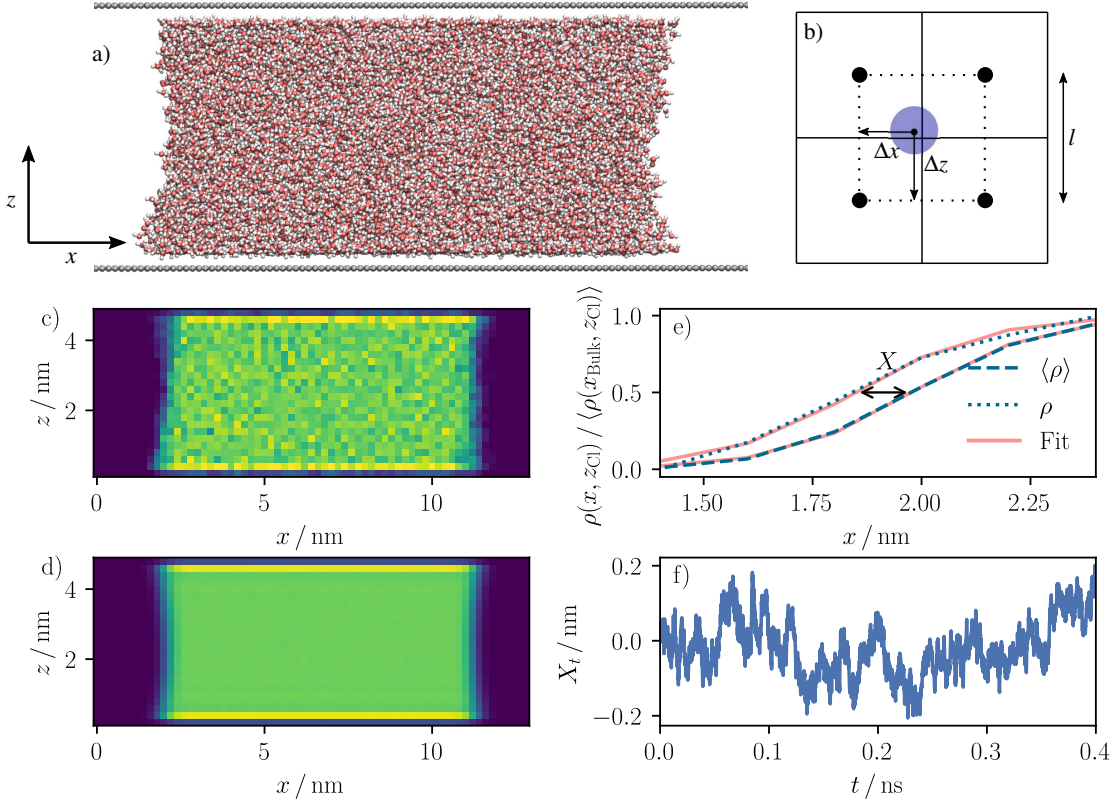
\begin{figure*}
    \centering
    \def\svgwidth{0.85\textwidth}
\begingroup%
  \makeatletter%
  \providecommand\color[2][]{%
    \errmessage{(Inkscape) Color is used for the text in Inkscape, but the package 'color.sty' is not loaded}%
    \renewcommand\color[2][]{}%
  }%
  \providecommand\transparent[1]{%
    \errmessage{(Inkscape) Transparency is used (non-zero) for the text in Inkscape, but the package 'transparent.sty' is not loaded}%
    \renewcommand\transparent[1]{}%
  }%
  \providecommand\rotatebox[2]{#2}%
  \newcommand*\fsize{\dimexpr\f@size pt\relax}%
  \newcommand*\lineheight[1]{\fontsize{\fsize}{#1\fsize}\selectfont}%
  \ifx\svgwidth\undefined%
    \setlength{\unitlength}{495bp}%
    \ifx\svgscale\undefined%
      \relax%
    \else%
      \setlength{\unitlength}{\unitlength * \real{\svgscale}}%
    \fi%
  \else%
    \setlength{\unitlength}{\svgwidth}%
  \fi%
  \global\let\svgwidth\undefined%
  \global\let\svgscale\undefined%
  \makeatother%
  \begin{picture}(1,0.71212124)%
    \lineheight{1}%
    \setlength\tabcolsep{0pt}%
    \put(0,0){\includegraphics[width=\unitlength]{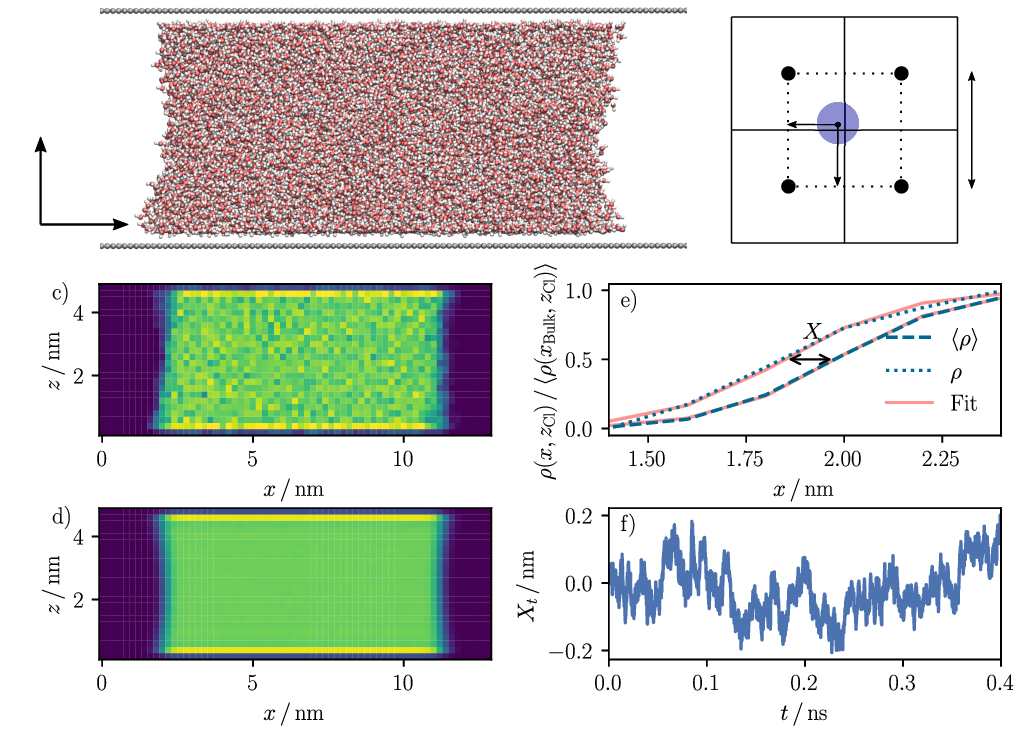}}%
    \put(0.07208691,0.47767875){\color[rgb]{0,0,0}\makebox(0,0)[lt]{\lineheight{1.25}\smash{\begin{tabular}[t]{l}$x$\end{tabular}}}}%
    \put(0.02001714,0.52671816){\color[rgb]{0,0,0}\makebox(0,0)[lt]{\lineheight{1.25}\smash{\begin{tabular}[t]{l}$z$\end{tabular}}}}%
    \put(0.9466949,0.58029539){\color[rgb]{0,0,0}\makebox(0,0)[lt]{\lineheight{1.25}\smash{\begin{tabular}[t]{l}$l$\end{tabular}}}}%
    \put(0.0948933,0.6540671){\makebox(0,0)[lt]{\lineheight{1.25}\smash{\begin{tabular}[t]{l}a)\end{tabular}}}}%
    \put(0.71573526,0.6728321){\makebox(0,0)[lt]{\lineheight{1.25}\smash{\begin{tabular}[t]{l}b)\end{tabular}}}}%
    \put(0.82085932,0.55746786){\makebox(0,0)[lt]{\lineheight{1.25}\smash{\begin{tabular}[t]{l}$\Delta z$\end{tabular}}}}%
    \put(0.77736234,0.56866927){\makebox(0,0)[lt]{\lineheight{1.25}\smash{\begin{tabular}[t]{l}$\Delta x$\end{tabular}}}}%
  \end{picture}%
\endgroup%

    \caption{
    (a) A simulation snapshot of the water–graphene liquid bridge viewed in the $xz$-plane.
    (b) A schematic illustration of how the density field is calculated to obtain a continuously changing field in time.
    The black dots indicate the center of each grid cell with cell length $l$ and $\Delta x$ and $\Delta z$ the distance of the particle to the lower left cell along the $x$- and $z$-axis respectively.
    The contribution of a particle (blue circle) to the density field is distributed to the four nearest grid cells via bilinear interpolation, e.g., the contribution to the bottom left cell is given by $(l-\Delta x)(l-\Delta z)/l^2$.
    (c) The instantaneous density profile $\rho(x, z)$ in the $xz$-plane and (d) its ensemble average $\expval{\rho(x, z)}$.
    (e) The density profile along $x$ at $z_{\mathrm{Cl}}$ for the instantaneous density (dotted blue line) and its time average (dashed blue line). Both profiles are approximated with Eq.~\ref{eq:fit} (solid red lines) to determine the contact line position from the interfacial position.
    (f) The resulting trajectory of the contact line position $X$. 
    }
    \label{fig:systems}
\end{figure*}
We review a derivation of Eq.~\ref{eq:gle} in supplementary information (SI) S1.
We denote functions of time with brackets, e.g., $K(t)$, and the time dependence of observables along a microscopic trajectory by subscript $t$, e.g., $V_t$.
We include a harmonic force in Eq.~\ref{eq:gle}, because if the droplet is at rest, the contact line will fluctuate around some average position $\expval{X}=0$.
While the memory kernel can be obtained from projected force correlation functions,\cite{carof_two_2014} this is usually not done.
Multiplying the Eq.~\ref{eq:gle} with $V_0$ yields a Volterra equation
\begin{equation}
    m\dot{C}(t) = -k\expval{V_0, X_t} -\int_0^{\infty}\dd s\, K(s)C(t-s), \label{Eq:volterra}
\end{equation}
where $\expval{\dots}$ denotes an ensemble average and $C(t) = \expval{V_0, V_t}$ is the velocity auto-correlation function (VACF).
This equation can be inverted\cite{straube_rapid_2020, kowalik_memory-kernel_2019} to extract the memory kernel from the corresponding VACF.
We want to point out that Eq.~\ref{eq:gle} is exact if the probability distribution of the contact line position around its average is Gaussian, i.e., the potential of mean force is harmonic.

For a stationary non-equilibrium ensemble under an external perturbation $P$, it follows that\cite{kubo1991}
\begin{equation}
    m\frac{\dd}{\dd t} \expval{V}_1 (t) = P(t) - \int^{\infty}_{-\infty}\dd s\, K_+(t-s)\expval{V}_1(s),
\end{equation}
where the subscript $+$ indicates multiplication with the unity step function and  $\expval{\dots}_1$ denotes the average over the non-equilibrium ensemble.
Fourier transforming this expression and rearranging for the velocity yields
\begin{equation}
    \expval{\hat{V}}_1(\omega) = \frac{1}{-i\omega m+\hat{K}_+(\omega)}\hat{P}(\omega) = \hat{\Upsilon}(\omega)\hat{P}(\omega), \label{eq:response}
\end{equation}
where $\hat{\Upsilon}(\omega)$ is the (linear) response function of the contact line. 
Here $\Re(\hat{\Upsilon}(\omega))$ is the dissipative response of the contact line and $\Im(\hat{\Upsilon}(\omega))$ is the elastic response.
The frequency-dependent contact line friction  $\zeta$ follows from the dissipative response
\begin{equation}
    \zeta(\omega) = \frac{1}{L}\Re(\hat{\Upsilon}^{-1}(\omega)) = \frac{1}{L} \Re(\hat{K}_+(\omega)),
\end{equation}
where $L$ is the length of the contact line.
Computing the memory kernel via Eq.~\ref{Eq:volterra} usually requires correlation functions with a high time resolution, but the zero-frequency contact line friction coefficient can be obtained from the position correlation function at a much lower resolution with a Green-Kubo-like relation\cite{kowalik_memory-kernel_2019}
\begin{equation}
     \zeta(0) = \frac{k_{\mathrm{B}} T}{L} \int^{\infty}_0\dd t\, \frac{\expval{X_0,X_t}}{\expval{X_0, X_0}^2}, \label{eq:green_kubo}
\end{equation}
where $k_{\mathrm{B}}$ is the Boltzmann constant and $T$ is the temperature.
We review a derivation of Eq.~\ref{eq:green_kubo} in SI~S2.

\section{Extracting the Contact Line Position} \label{sec:def_cl}
\begin{figure*}
    \centering
    \includegraphics[width=0.9\linewidth]{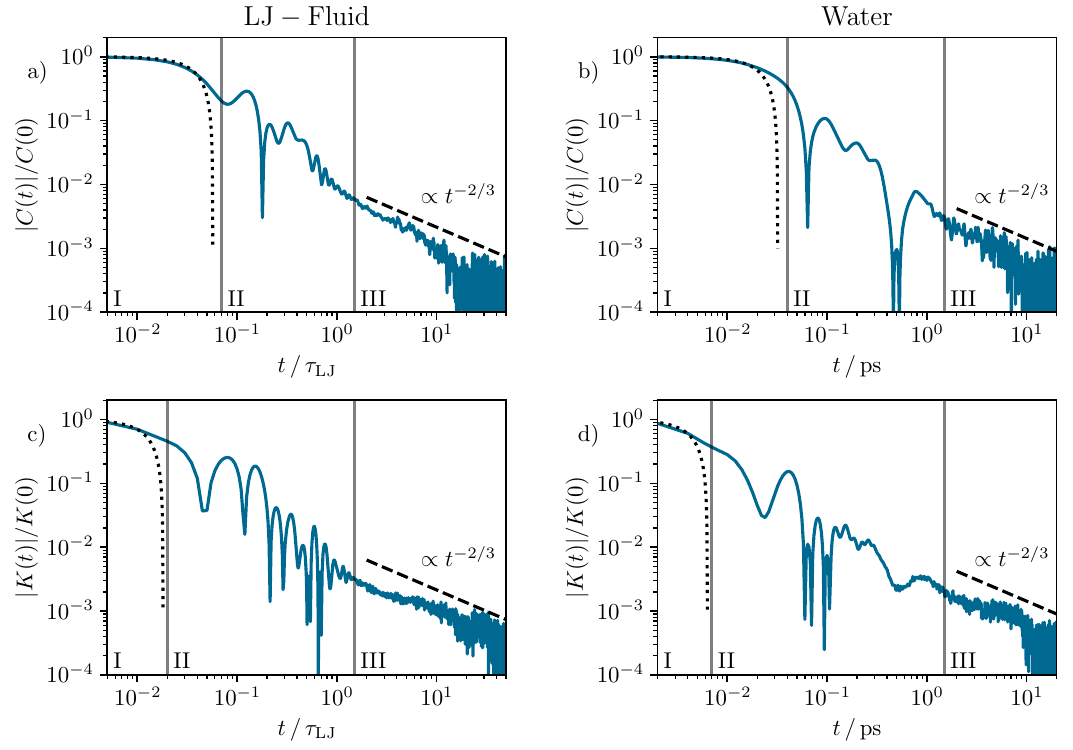}
    \caption{
    (a) the VACF and (c) the corresponding memory kernel $K$ for an LJ-octamer fluid confined between generic solid plates, and (b) the VACF and (d) the corresponding memory kernel for water confined between graphene sheets. 
    For both systems, the VACF and, to a lesser extent, the memory kernel are well captured by a short-time expansion of the form $1 - (t\omega_0)^2$ (dotted black lines) with the slope approaching zero for $t\rightarrow 0$.
    This indicates that the contact line velocity is a well-defined observable.
    At long times, both the VACF and memory kernel decay according to a power law (dashed black lines) with an exponent of $2/3$.
    }
    \label{fig:cfs}
\end{figure*}

We study two liquid bridge systems: A generic Lennard-Jones (LJ) fluid, where each molecule consists of eight beads, with negligible vapor pressure confined between two simple cubic plates with a contact angle of 113.0\textdegree$\pm$0.2\textdegree, and water confined between graphene sheets with a contact angle of 80.3\textdegree$\pm$0.6\textdegree, shown in Fig.\ref{fig:systems} (a). 
Details on the molecular dynamics (MD) simulations are provided in the Appendix and details on the contact angle extraction in SI~S3.
Projecting the system onto the $xz$-plane reveals four contact lines aligned along the $y$-axis. 
To determine their positions, we compute the density field $\rho(x,z)$ on a grid with cell length $l$.
We then follow common approaches used in, e.g., hybrid particle–field theory \cite{milano_hybrid_2009} to distribute the contribution of each particle to the surrounding cells by bilinear interpolation, schematically shown in \ref{fig:systems} (b).
This ensures that the contact line position changes continuously in time (see SI~S4).
The resulting density profile and its ensemble average are shown in Fig.~\ref{fig:systems} (c) and (d), respectively.
We assume the $z$-position of the contact line $z_{\mathrm{Cl}}$ to be at the first density maximum normal to the plate and extract the position along $x$ from the density profile $\rho(x,z_{\mathrm{Cl}})$ (dotted blue line) and its ensemble average (dashed blue line) shown in Fig.~\ref{fig:systems} (e).
Using that both systems have a negligible vapor pressure, we approximate the density profiles (solid red lines) with,
\begin{equation}
\frac{\rho(x,z_{\mathrm{Cl}})}{\langle \rho(x_{\mathrm{L}},z_{\mathrm{Cl}}) \rangle}
= \frac{1}{2} + \frac{1}{2}\tanh(2\frac{x-x_{\mathrm{Cl}}}{w}),
\label{eq:fit}
\end{equation}
where $w$ is the interfacial width and $x_{\mathrm{L}}$ denotes a position where the density at the solid-liquid interface is not influenced by the liquid-gas interface.
Finally, we obtain the contact line position as the difference between the instantaneous and the average interfacial position.
This procedure yields a smooth contact line trajectory shown in Fig.~\ref{fig:systems} (f), with a Gaussian probability density for both the LJ- and the water–graphene system (see SI~S5), from which the contact line velocity is obtained by numerical differentiation.
Finite size effects with respect to the length of the contact line do not play a major role (see SI~S6).

\section{Non-Markovian Effects and Frequency Dependent Response} \label{sec:cl_dynamics}
\begin{figure*}
    \centering
    \includegraphics[width=0.9\linewidth]{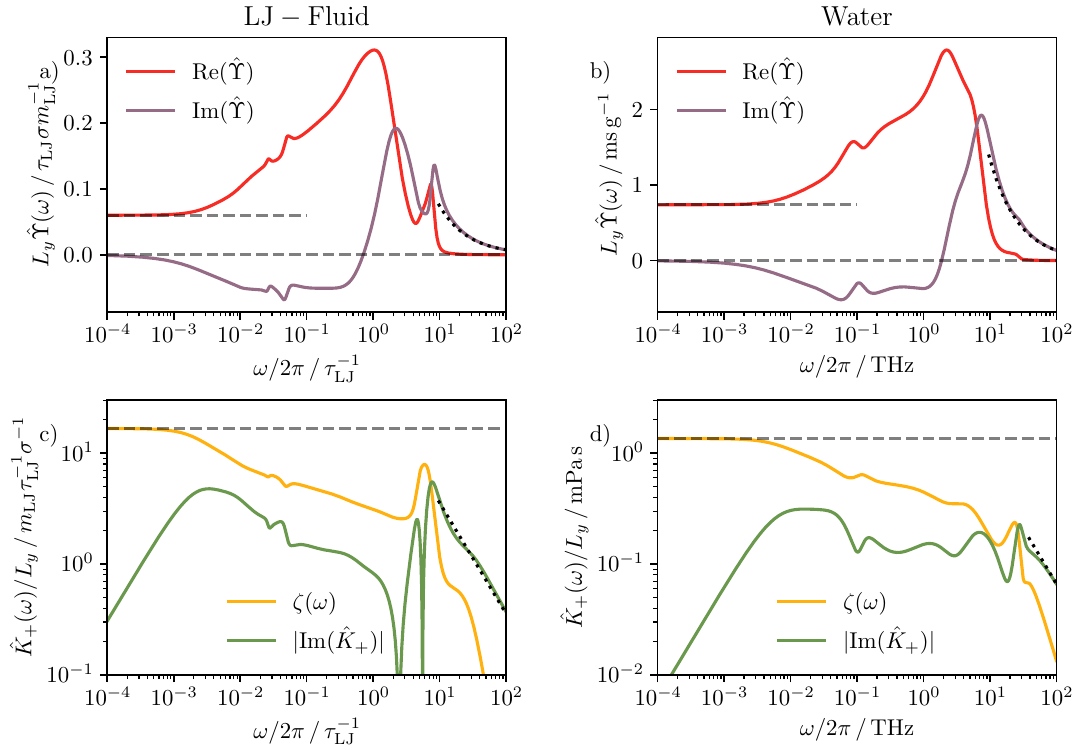}
    \caption{
    The dissipative $\Re(\hat{\Upsilon})$ (solid red lines) and elastic $\Im(\hat{\Upsilon})$ (solid purple lines) response of the contact line of a) an LJ-octamer fluid confined between generic solid plates, and b) water confined between graphene sheets. 
    The real part of the half-sided Fourier transform of the memory kernel $\Re(\hat{K}_{+})$, corresponding to the frequency-dependent contact line friction $\zeta(\omega)$ (solid yellow lines), and the imaginary part of the transform $\Im(\hat{K}_{+})$  of c) an LJ-octamer fluid confined between generic solid plates, and d) water confined between graphene sheets. 
    }
    \label{fig:fricwL}
\end{figure*}

Figure~\ref{fig:cfs} shows the VACF $C(t)$ and the corresponding memory kernel $K(t)$, obtained by inverting Eq.~\ref{Eq:volterra}, on a logarithmic scale for (a,c) a generic LJ liquid bridge and (b,d) a water–graphene liquid bridge.
We can distinguish three regimes labeled I-III.
At short times (regime I), the VACF -and to a lesser extent the memory kernel- follow a quadratic expansion  $1 -  \nu^2t^2 + \mathcal{O}(t^4)$ (black dotted lines).
Odd-order derivatives vanish due to the skew symmetry of time derivatives.
The excellent agreement of the VACF with the quadratic expansion confirms that the contact line velocity is a well-defined observable.
By contrast, $K(t)$ deviates more strongly, which we attribute to numerical limitations.
Following the initial decay, both the VACF and the memory kernel show damped oscillations (regime II).
We shown in SI~S7, that these oscillations coincide with low–frequency molecular vibrations arising from fast, molecular-scale motions.
For the water–graphene contact line, in particular, we observe a strong peak at $1.2$\,THz in the Fourier transform of the VACF coinciding with SPC/Es low-frequency librations.\cite{carey_measurement_1998,lin_two-phase_2010}
At large times (regime III), both the VACF and the memory kernel follow a power law decay with an exponent of $-2/3$.
The shared asymptotic behavior follows from the Laplace transform of  Eq.~2.\cite{corngold_behavior_1972}
Power-law decays in correlation functions are generally a result of coupling to slow modes of the system.
Note that if the movement of the contact line were determined by fast processes, the decay of correlations would be exponential.
We show in SI~S8 that all four contact lines couple weakly to each other, indicating that the contact lines couple to hydrodynamic modes.
This coupling results in corresponding finite size effects (see SI~S8).

Fig~\ref{fig:fricwL} shows the dissipative response $\Re(\hat\Upsilon)$ (solid red line) and the elastic response $\Im(\hat\Upsilon)$ (solid purple line) given by Eq.~\ref{eq:response} for (a) the LJ-fluid and (b) the water-graphene liquid bridge.
For further details on the computation of the Fourier transform, see SI~S9.
In close analogy to hydrodynamic friction, at low frequencies, the response is static and dissipative.
In this case, the response is well described by a Markovian contact line friction coefficient $L_y \hat\Upsilon(\omega\to 0) = 1/\zeta(0)$ (upper gray dashed line), and there is no elastic response.
At high frequencies beyond the characteristic frequency of the contact line, corresponding to a maximum in the elastic response, dissipation decreases exponentially to zero (lower gray dashed line), and the contact line behaves elastically. 
In this limit the response follows the asymptotic $L_y\hat\Upsilon(\omega\to\infty) =  i L_y/(m\omega)+\mathcal{O}(\omega^{-3})$ (black dotted line).
Fig.~\ref{fig:fricwL} shows the frequency-dependent contact line friction $\zeta(\omega) = \Re(\hat{K}_{+}(\omega))$ (yellow solid line) and the corresponding imaginary part $\Im(\hat{K}_{+})$ (solid green line) for (c) the LJ-fluid and (d) water-graphene liquid bridge.
At low frequencies, the contact line friction approaches the above-mentioned Markovian contact line friction coefficient, followed by a decrease in friction. 
The time scales in this frequency range fall into regime III of the VACF and memory kernel, meaning the majority of the contact-line friction emerges on the same time scales where the contact line couples to slow modes.
At high frequencies, beyond the characteristic frequency of the system, the contact line friction decreases since the real part of $\hat\Upsilon$ vanishes, with the imaginary part following an asymptotic $\Im(\hat{K}_{+}(\omega\to\infty))=iK(0)/\omega+\mathcal{O}(\omega^{-3})$ (dotted black line).

While MKT correctly captures the relationship between the work of adhesion and contact-line friction,\cite{blake_influence_2002, de_ruijter_dynamic_1999, duvivier_experimental_2011, nold_hydrodynamic_2024} the picture that contact-line motion is driven by fast absorption processes involving only single molecules is not supported by our findings.
While at intermediate times (Regime II in Fig.~\ref{fig:cfs}), the contact-line motion is dominated by molecular vibrations, these oscillations are not followed by an exponential decay of correlations at large times (Regime III in Fig.~\ref{fig:cfs}).
In contrast, we find that at large times the dynamics are governed by coupling to the hydrodynamics of the system with the majority of the contact-line friction emerging due to this coupling.
This observation is in conflict with descriptions of dynamic wetting where the movement of the contact line is attributed to rapid processes involving only a few molecules at the contact line, like MKT.

\section{Conclusion} \label{sec:conc}
We have derived a generalized description of contact-line friction, in close analogy to the description of hydrodynamic friction.
This approach yields, within the linear response regime, a frequency-dependent description of contact-line friction obtained directly from equilibrium data.
We introduce a robust procedure to extract the contact-line coordinate and its correlation functions from MD data, and show that the corresponding correlation functions obey the expected short-time expansion of well-defined variables.
The VACF and memory kernel feature damped oscillations caused by low-frequency molecular vibrations, and on large time scales a power law decay arising from coupling to the systems hydrodynamic modes.
From the frequency-dependent friction, we infer that friction mostly emerges due to this coupling.
As a result the microscopic contact line dynamics are neither Markovian nor determined by the movement of a few molecules at the contact line.
On the contrary the dynamics are strongly non-Markovian and influenced by hydrodynamic modes in the bulk of the droplet.

\subsection*{Acknowledgments}
The authors thank Vishal Dadich, Viktor Klippenstein, Florian Müller-Plathe, and Burkhard Dünweg for helpful discussions.
This project was funded by the Deutsche Forschungsgemeinschaft (DFG, German Research Foundation) in the framework to the collaborative research center “Multiscale Simulation Methods for Soft Matter Systems” (TRR 146) under Project No. 233630050. 

\subsection*{Data Availability Statement}
The data that supports the findings of this study, input files for simulations, and the raw data used to generate the figures shown is freely \href{https://tudatalib.ulb.tu-darmstadt.de/handle/tudatalib/4903}{available}.

\bibliography{bib_cor}

\appendix

\section{MD Simulation Details}\label{app:md_details_sim}
All MD simulations are performed with LAMMPS\cite{thompson_lammps_2022} (stable release 2. Apr. 2025).
Integration of the liquid particles is performed in the $NVT$ ensemble. 
The positions of the solid particles are not updated.
The center of mass of the liquid is fixed by removing the center of mass momentum of the liquid at every step.
Initial configurations are generated using the molecular builder Moltemplate\cite{jewett_moltemplate_2021}, and snapshots visualized with VMD.\cite{humphrey_vmd_1996}
During production runs, the density field is computed with a custom LAMMPS compute.
From the density, we obtain the contact line position as outlined in the main text. 
Correlation functions are obtained via fast Fourier transforms as implemented in NumPy\cite{harris_array_2020} and then averaged over $10$ production runs and the $4$ contact lines.

For the generic LJ liquid, the mass $m_{\mathrm{LJ}}$, the depth of the potential well $\epsilon_{\mathrm{LL}}$ and the distance $\sigma$ at which the potential is $0$ are set to $1$.
Each molecule is modeled as an eight-bead oligomer with adjacent beads connected by finitely extensible non-linear elastic bonds\cite{kremer_dynamics_1990} using a bond constant of $k=35\,\epsilon\,\sigma^{-2}$ and an extensibility of $R_0=1.5\,\sigma$ and all interactions are cutoff at $2.5\,\sigma$.
The thermal energy $k_{\mathrm{B}}T$ is equal $\epsilon_{\mathrm{LL}}$ and the integration time step is  $0.005\,\tau_{LJ}$ for all simulations.
Both solid plates are modeled as single-layer primitive cubic sheets with a lattice constant of $0.5\,\sigma$.
Liquid-solid interactions are modeled with LJ-parameters of $\epsilon_{\mathrm{LS}} = 0.125\,k_{\mathrm{B}}T$ and $\sigma_{\mathrm{LS}}=1$.
The liquid phase contains $4,000$ molecules, for a total of $32,000$ liquid beads, and the two solid sheets contain $16,000$ particles each, for a total of $32,000$ solid particles in a simulation box with dimensions $L_x=100\,\sigma$, $L_y=40\,\sigma$, and $L_z=21.001\,\sigma$, with a distance of $21.0\,\sigma$ between the solid plates.
The boundaries in the $x$ and $y$ directions are periodic, and a non-periodic boundary in the $z$ direction.
To relax the liquid density, we first integrate the system for $2,500\,\tau_{\mathrm{LJ}}$  with a Langevin thermostat\cite{schneider_molecular-dynamics_1978,dunweg_brownian_1991} constant of $0.5\,\tau_{\mathrm{LJ}}$.
Then the contact angle is equilibrated for $25,000\,\tau_{\mathrm{LJ}}$ using a Nosé–Hoover-thermostat\cite{tuckerman_liouville-operator_2006} with a thermostat constant of $1\,\tau_{\mathrm{LJ}}$. 
To obtain $10$ independent configurations, we randomize the velocity of all liquid particles and integrate the system using a Nosé–Hoover thermostat with a thermostat constant of $5\,\tau_{\mathrm{LJ}}$ for $50,000\,\tau_{\mathrm{LJ}}$.
We then perform a production run with all $10$ systems using a Nosé–Hoover thermostat with a thermostat constant of $5\,\tau_{\mathrm{LJ}}$ for $50,000\,\tau_{\mathrm{LJ}}$, outputting the density field at every step.
The density field is calculated from the bead positions (not the center of mass).

Water molecules are modeled with the SPC/E water model.\cite{berendsen_missing_1987}
Water–graphene interactions are described by a Lennard-Jones potential with LJ-parameters $\epsilon_{\mathrm{WC}}=0.4389\,\mathrm{kJ}\,\mathrm{mol}^{-1}$ and $\sigma_{\mathrm{WC}}=0.3367\,\mathrm{nm}$.\cite{carlson_modeling_2024}
Bond lengths and angles in water are constrained to their equilibrium values using the SHAKE algorithm with a tolerance of $0.0001$ and a maximum of $10$ iterations.\cite{ryckaert_numerical_1977} 
The system consists of $14,850$ water molecules and $20,000$ carbon atoms in two graphene sheets with a $CC$ bond length of $0.142\,\mathrm{nm}$.
The dimensions of simulation box are $L_x=24.595\mathrm{nm}$, $L_x=10.65\mathrm{nm}$, and $L_z=5.0001\mathrm{nm}$ with a distance of $5.0\,\mathrm{nm}$ between the graphene sheets.
The boundary is periodic in the $x$- and $y$-directions and non-periodic in the $z$-direction. 
Long-range electrostatics are treated with the slab particle–particle particle–mesh method using a real-space cutoff of $1.0\,\mathrm{nm}$, an accuracy of $10^{-4}$, and a slab factor of $3$.
We first minimize the energy of the system and then equilibrate for $4\,\mathrm{ns}$ using a Nosé–Hoover thermostat with a thermostat constant of $1\,\mathrm{ps}$.
To obtain $10$ independent configurations, we randomize the velocity of all liquid particles and then integrate the system for $2\,\mathrm{ns}$ using a Nosé–Hoover thermostat with a thermostat constant of $1\,\mathrm{ps}$.
Finally, we perform production runs with all $10$ systems for $10\,\mathrm{ns}$ using a Nosé–Hoover thermostat with a thermostat constant of $1\,\mathrm{ps}$, outputting the density field every step.
The density field is calculated from the position of the oxygen atoms.


\end{document}


\title{Supplemental Material for "Memory Effects in Contact Line Friction"}
\author{Niklas Wolf, Nico F. A. van der Vegt \\
\textit{Department of Chemistry, Technical University of Darmstadt, 64287 Darmstadt, Germany}}
\date{}

\maketitle

\section{Additional information on the GLE}
In the following, we will briefly review a derivation of Eq.~1.
Consider a Hamiltonian system of $N$ particles with positions $\bm{q}^N$ and momenta $\bm{p}^N$ in some volume $\Omega$ with temperature $T$.
The Hamiltonian of the system is given by
\begin{equation}
    H = \bm{p}^N \cdot\frac{M^{-1}}{2}\bm{p}^N + U(\bm{q}^N), 
\end{equation}
where $\bm{M}$ is the mass matrix of the system.
In equilibrium, the probability density of finding a configuration $(\bm{q}^N,\bm{p}^N)$ is given by the Boltzmann factor
\begin{equation}
    f(\bm{q}^N,\bm{p}^N) = \frac{e^{-\beta H(\bm{q}^N,\bm{p}^N)}}{Q},
\end{equation}
where $Q$ is the partition function normalizing the density and $\beta = \frac{1}{k_{\mathrm{B}}T}$ is the inverse temperature.
We are interested in the time evolution of two specific observables, the deviation of the contact line position $X(\bm{q}^N)$ from its average position and its time derivative $V(\bm{q}^N, \bm{p}^N) = \frac{\dd X}{\dd t}$.
The observables of the system, as in any function of $\bm{q}^N$ and $\bm{p}^N$, are members of a vector space where the scalar product is given by the ensemble average.
For example, the scalar product between the contact line position and its velocity is
\begin{equation}
    \expval{X, V} = \int_{\Omega^N} \dd \bm{q}^N \int_{\mathbb{R}^{3N}} \dd\bm{p}^N \,f(\bm{q}^N,\bm{p}^N) X(\bm{q}^N)V(\bm{q}^N,\bm{p}^N).
\end{equation}
Further the time evolution of any observable is formally given by the Liouville equation
\begin{align}
    \frac{\dd V_t}{\dd t} =& \ \mathcal{L}V_t
    = \bm{p}^N_t \cdot \bm{M}^{-1}\partial_{\bm{q}^N}V_t-\partial_{\bm{q}^N}U(\bm{q}^N_t)\cdot\partial_{\bm{p}^N}V_t, \label{eq:liou_v}
\end{align}
where $\mathcal{L}$ is the Liovillian or Liouville operator.
The Liouvillian is skew-symmetric with respect to the scalar product, meaning the scalar product of $V$ and $X$ may be written as
\begin{equation}
    \expval{V, X} = \expval{\mathcal{L}X, X} = - \expval{X, \mathcal{L}X} = -\expval{X, V} = 0.
\end{equation}
Integrating the Liouville equation gives a formal solution for the time evolution
\begin{equation}
    V_t = e^{t\mathcal{L}}V_0 \label{eq:v_t}
\end{equation}
in terms of a propagator $e^{t\mathcal{L}}$.

Within the Mori-Zwanzig formalism, the equation of motion is split with a projection into a part that depends on $X$ and $V$ and an orthogonal part.
The corresponding projection operator or projector is given by
\begin{equation}
    \mathcal{P} = \frac{\expval{X_0, \cdot}}{\expval{X_0, X_0}}X_0 + \frac{\expval{V_0, \cdot}}{\expval{V_0, V_0}}V_0.
\end{equation}
Splitting Eq.~\ref{eq:v_t} with this projector, we obtain
\begin{equation}
    \frac{\dd V_t}{\dd t} = e^{t\mathcal{L}}\mathcal{P}\mathcal{L}V_0 + e^{t\mathcal{L}}(1-\mathcal{P})\mathcal{L}V_0, \label{eq:gle_start}
\end{equation}
where the first term depends on $X_0$ and $V_0$, and the second term contains all other contributions to the equation of motion.
Using the skew symmetry of the Liouvillian, the first term in Eq.~\ref{eq:gle_start} can be evaluated to
\begin{equation}
    e^{t\mathcal{L}}\mathcal{P}\mathcal{L}V_0 = e^{t\mathcal{L}}\frac{\expval{X_0, \mathcal{L}V_0}}{\expval{X_0, X_0}}X_0 + e^{t\mathcal{L}}\frac{\expval{V_0, \mathcal{L}V_0}}{\expval{V_0, V_0}}V_0 =  -e^{t\mathcal{L}}\frac{\expval{\mathcal{L}X_0, V_0}}{\expval{X_0, X_0}}X_0 = -\frac{\expval{V_0, V_0}}{\expval{X_0, X_0}} X_t.
\end{equation}
The variance of the velocity is related to the contact line mass $m$ via the equipartition theorem
\begin{equation}
    \expval{V_0, V_0} = \frac{1}{m\beta}
\end{equation}
and the variance of the position is related to a (harmonic) force constant
\begin{equation}
    \expval{X_0, X_0} = \frac{1}{k\beta}.
\end{equation}
The first term reduces in Eq.~\ref{eq:gle_start} to
\begin{equation}
   -\frac{\expval{V_0, V_0}}{\expval{X_0, X_0}} X_t = -\frac{k}{m}X_t.
\end{equation}
To evaluate the second term in Eq.~\ref{eq:gle_start}, we decompose the propagator with the projection operator using the Duhamel or Dyson formula 
\begin{equation}
    e^{t\mathcal{L}} = \int^t_0\dd s\, e^{(t-s)\mathcal{L}}\mathcal{P} \mathcal{L}e^{s(1-\mathcal{P})\mathcal{L}} + e^{t(1-\mathcal{P})\mathcal{L}}.
\end{equation}
Applying this to the second term in Eq.~\ref{eq:gle_start} gives
\begin{equation}
     e^{t\mathcal{L}}(1-\mathcal{P})\mathcal{L}V_0 = \int^t_0\dd s\, e^{(t-s)\mathcal{L}}\mathcal{P} \mathcal{L}\frac{F^R_s}{m} + \frac{F^R_t}{m},
\end{equation}
where we introduced the random or stochastic force 
\begin{equation}
    \frac{F^R_t}{m} =  e^{t(1-\mathcal{P})\mathcal{L}} (1-\mathcal{P})\mathcal{L}V_0.
\end{equation}
Using the skew symmetry of the Liouvillian, the integral term can be evaluated to
\begin{equation}
     \int^t_0 \dd s\, e^{(t-s)\mathcal{L}}\mathcal{P} \mathcal{L}\frac{F^R_s}{m}  = -\int^t_0\dd s\, e^{(t-s)\mathcal{L}}\frac{\expval{F_0, F^R_s}}{m{\expval{V_0, V_0}}}V_{0} - \int^t_0\dd s\,e^{(t-s)\mathcal{L}}\frac{\expval{V_0, F^R_s}}{m{\expval{X_0, X_0}}}X_{0}.
\end{equation}
The random force is, by construction, uncorrelated to the velocity at time zero, so the second term vanishes.
At $t=0$, the force is the sum of the harmonic and random force $F_0 = -k X_0 + F^R_0$.
Since the random force is, by construction, also uncorrelated to the position at time zero, we obtain
\begin{equation}
     \int^t_0 \dd s\, e^{(t-s)\mathcal{L}}\mathcal{P} \mathcal{L}\frac{F^R_s}{m} = -\int^t_0\dd s\,\frac{\expval{F^R_0, F^R_s}}{m^2{\expval{V_0, V_0}}}V_{t-s}.
\end{equation}
Introducing the memory kernel $K$ 
\begin{equation}
    K(s) = \frac{\expval{F^R_0, F^R_s}}{m{\expval{V_0, V_0}}},
\end{equation}
we obtain a GLE describing the motion of the contact line
\begin{equation}
    m \frac{\dd V}{\dd t} = F_t = -kX_t-\int_0^{t}\dd s\, K(s)V_{t-s} + F^R_t. \label{eq:gle}
\end{equation}
The equation is closed by the fluctuation-dissipation relation connecting the correlation function of $F^R$ to $K$
\begin{equation}
    K(t) =  \frac{\expval{F^R_0, F^R_t}}{m{\expval{V_0, V_0}}} = \beta \expval{F^R_0, F^R_t}. \label{eq:fdr}
\end{equation}
This derivation is exact in the sense that there are no approximations made.
We have simply restated the original equation of motion.
However, for most applications we require the statistics of $F^R$ to be Gaussian.
\clearpage

\section{Additional information on the derivation of Eq.~6}
In the following, we will briefly derive Eq.~6.
Extending the integral in Eq.~1 to negative infinity yields 
\begin{equation}
    m \frac{\dd V}{\dd t}= -kX_t-\int_{-\infty}^{\infty}\dd s\, K_{+}(s)V_{t-s} + F^R_t, \label{eq:gle_neg}
\end{equation}
where the subscript $+$ indicates multiplication by the unity step function. 
Fourier transforming 
\begin{equation}
     -m\omega^2 \hat{X}_{\omega} = -k\hat{X}_{\omega} - i\omega\hat{K}_{+}(\omega)\hat{X}_{\omega} + \hat{F}^R_\omega
\end{equation}
and rearranging for $\hat{X}_{\omega}$ gives 
\begin{equation}
    \hat{X}_{\omega} = \frac{1}{k + i\omega\hat{K}_{+}(\omega) -  m\omega^2}\hat{F}^R_\omega = \hat{\chi}(\omega)\hat{F}^R_\omega,\label{eq:x_resp}
\end{equation}
where we introduced the response function $\chi$. 
Multiplying with $X_0$ and inserting Eq.~\ref{eq:x_resp} yields
\begin{equation}
    \expval{X_0, \hat{X}_{\omega}} = \hat{\chi}(\omega)\expval{X_0, \hat{F}^R_{\omega}} = \int^{\infty}_{-\infty}\frac{\dd\omega'}{2\pi} \, \hat{\chi}(\omega)\expval{\hat{X}_{\omega'}, \hat{F}^R_{\omega}} = \int^{\infty}_{-\infty}\frac{\dd\omega'}{2\pi} \, \hat{\chi}(\omega)\hat{\chi}(\omega')\expval{\hat{F}^R_{\omega'}, \hat{F}^R_{\omega}}, \label{eq:ft_cxx}
\end{equation}
where we used that
\begin{equation}
    X_0 = \int^{\infty}_{-\infty} \frac{\dd\omega}{2\pi}\, \hat{X}_{\omega}.
\end{equation}
To substitute the inner product of the random force in Eq.~\ref{eq:ft_cxx}, we fourier transform Eq.~\ref{eq:fdr} 
\begin{align}
    \beta \expval{\hat{F}^R_{\omega'}, \hat{F}^R_\omega} = \int^{\infty}_{-\infty}\dd t'\, e^{i\omega' t'} \int^{\infty}_{-\infty}\dd t\, e^{i\omega t} K(t-t') 
    = \hat{K}(\omega)\int^{\infty}_{-\infty}\dd t'\, e^{i\omega' t'} e^{-i\omega t'}
    = 2\pi\delta(\omega'+\omega)\hat{K}(\omega)
\end{align}
Combining the Fourier transformed FDR with Eq.~\ref{eq:ft_cxx} gives
\begin{equation}
    \expval{X_0, \hat{X}_{\omega}} = k_{\mathrm{B}}T  \int^{\infty}_{-\infty}\dd\omega' \, \hat{\chi}(\omega)\hat{\chi}(\omega')\delta(\omega'+\omega)\hat{K}(\omega) = k_{\mathrm{B}}T  \hat{\chi}(\omega)\hat{\chi}(-\omega)\hat{K}(\omega).
\end{equation}
Rearranging for $\hat{K}(0)$, using the equipartition theorem, and the symmetry of correlation functions gives Eq.~6
\begin{equation}
    \zeta(0) = \frac{\hat{K}(0)}{2L} = \frac{1}{2k_{\mathrm{B}}TL\hat{\chi}(0)\hat{\chi}(0)}\int^{\infty}_{-\infty}\dd t\, \expval{X_0, X_t} = \frac{k_{\mathrm{B}}T}{L}\int^{\infty}_{0}\dd t\, \frac{\expval{X_0, X_t}}{\expval{X_0, X_0}^2}
\end{equation}
\clearpage


\section{Additional information on the calculation of Contact angles}
\begin{figure}[H]
    \centering
    \includegraphics{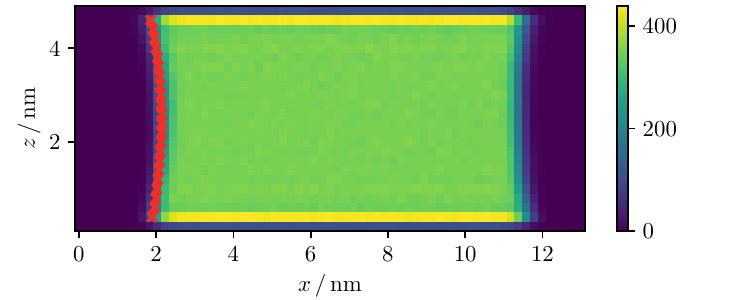}
    \caption{The averaged particle density profile with the miniscus on the left side indicated by crosses. 
    The density is given in units of nm$^{-3}$.
    The miniscus is fit with a circle to obtain a contact angle.}
    \label{fig:calc_cla}
\end{figure}

To calculate contact angles, we first extract the average interfacial position along the $x$-axis, $x_{\mathrm{I}}$, from the averaged density profile in the $xz$-plane $\expval{\rho(x, z)}$ via
\begin{equation}
        \expval{\rho(x_{\mathrm{I}}, z)} = \frac{\expval{\rho(x_{\mathrm{B}}, z)}}{2},\label{eq:interface}
\end{equation}
where $\rho(x_{\mathrm{B}})$ is a position in the bulk.
To increase numerical accuracy, we interpolate linearly between grid points.
Fig.~\ref{fig:calc_cla} shows the averaged density profile of water on graphene and the miniscus determined with Eq.~\ref{eq:interface}.
We fit a circle to the miniscus and calculate the contact angle via
\begin{equation*}
    \theta = \pi \pm \sin^{-1}\left(\frac{\left|z_{\mathrm{Cl}}-\frac{h}{2}\right|}{r}\right), 
\end{equation*}
where $z_{\mathrm{Cl}}$ is the $z$-coordinate of the contact line, $h$ is the distance between the plates, and $r$ is the radius of the circle fitted to the miniscus.
The sign depends on whether the miniscus is convex or concave.
\FloatBarrier
\clearpage

\section{Additional information on the Calculation of Density}
\begin{figure}[h]
    \centering
    \includegraphics[width=0.5\linewidth]{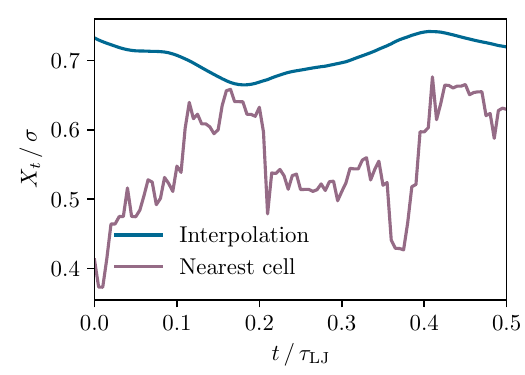}
    \caption{
    The contact line position obtained from a density that changes continuously in time (solid blue line) and the contact line position obtained from a density that does not change continuously in time (solid purple line).
    }
    \label{fig:bad_cl}
\end{figure}

Fig.~\ref{fig:bad_cl} shows the contact line trajectory obtained when the density is computed by counting the number of particles in each grid cell (purple line), and the contact line trajectory obtained when the density is calculated using bilinear interpolation (blue line).
We only obtain a continuous contact line trajectory when the density changes continuously in time.
\FloatBarrier
\clearpage


\section{Additional information on the Probability density of the Position and the Mass}
\begin{figure}[h]
    \centering
    \includegraphics[width=\linewidth]{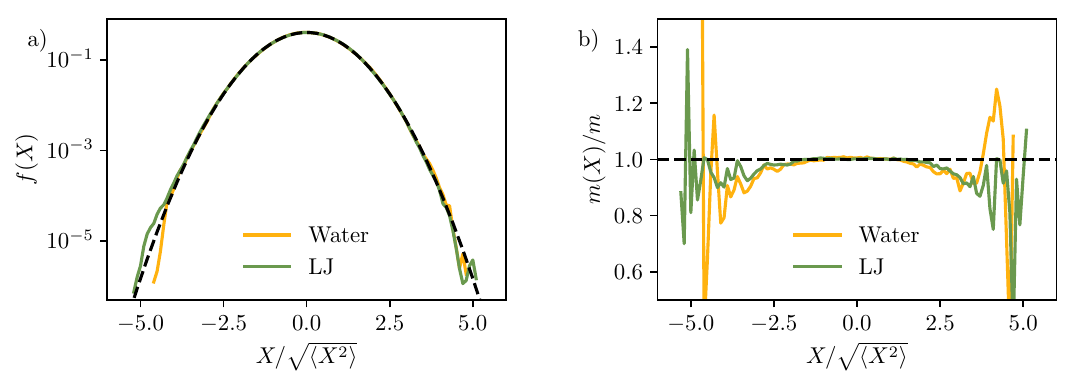}
    \caption{
    (a) The probability density $f$ of the contact line as a function of the contact line position normalized by the standard deviation of the contact line position.
    Both the water system (solid yellow line)  and the LJ system (solid green line) are in excellent agreement with a Gaussian distribution with zero mean and unity variance (dashed black line).
    (b) The position-dependent mass obtained from the position-dependent equipartition theorem Eq.~\ref{eq:pos_equi} as a function of the contact line position normalized by the standard deviation of the contact line position.
    For small fluctuations of the contact line position around equilibrium, the contact line mass of both the water system (solid yellow line) and the LJ system (solid green line) approaches the average value (dashed black line).
    }
    \label{fig:dens_pos_vel}
\end{figure}

Fig.~\ref{fig:dens_pos_vel}(a) shows the probability density $f$ of the contact line position for water (solid yellow line) and the LJ system (solid green line) as a function of the contact line position divided by the standard deviation of the contact line position. 
The dashed black line indicates a Gaussian distribution with zero mean and unity variance in excellent agreement with the probability density of both systems.
Fig.~\ref{fig:dens_pos_vel}(b) shows the position-dependent mass of the contact line position obtained from the position-dependent equipartition theorem
\begin{equation}
    m(X') = \frac{k_{\mathrm{B}}T}{\expval{V^2 \delta(X-X')}} \label{eq:pos_equi}
\end{equation}
for water (solid yellow line) and the LJ system (solid green line) as a function of the contact line position normalized by the standard deviation of the respective contact line position. 
The dashed black line indicates the average mass 
\begin{equation}
    m = \int^{\infty}_{-\infty}\dd X\, m(X) f(X) = \frac{k_{\mathrm{B}}T}{\expval{V^2}}.
\end{equation}
We find for both systems that around the equilibrium position of $X=0$, the mass is only very weakly position dependent and decreasing for large deviations from equilibrium.
Note that the fluctuations for very large deviations are a result of poor sampling of rare configurations far from the average position.
This indicates that Eq.~1 provides an excellent description of the contact line dynamics.
\FloatBarrier
\clearpage

\section{Additional information on Finite Size Effects}
\begin{figure}[h]
    \includegraphics[width=\linewidth]{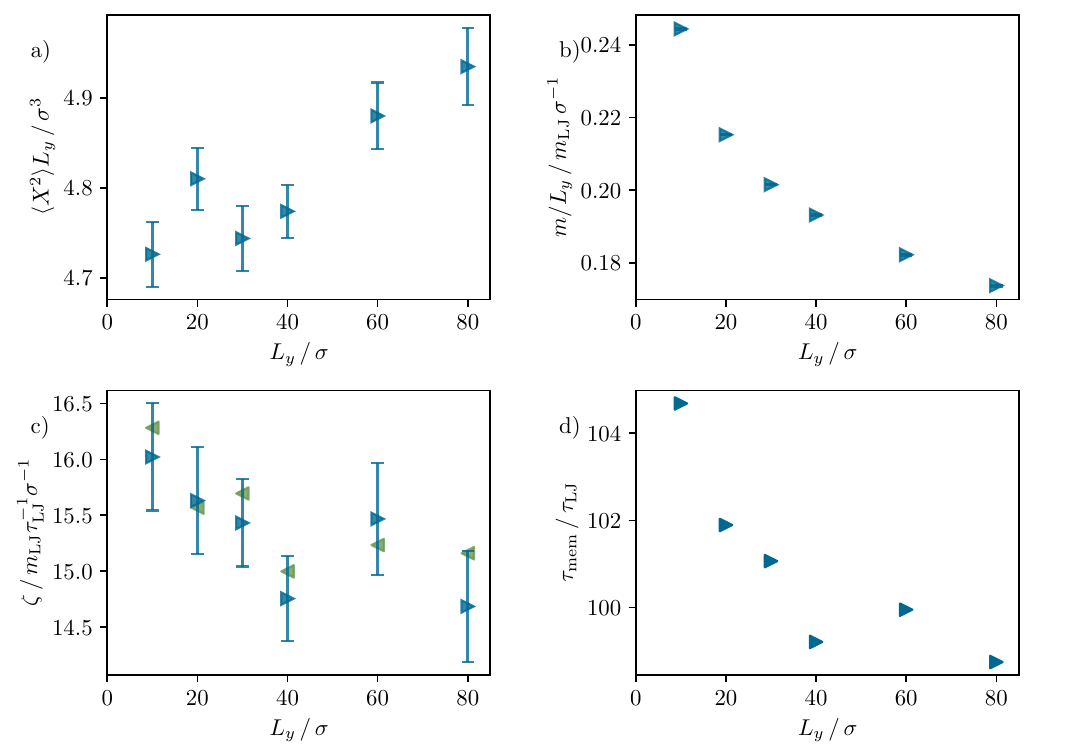}
    \centering
    \caption{
    (a) the mean amplitude of the contact line fluctuations multiplied by the length of the contact line.
    (b) The mass of the contact line divided by the length of the contact line.
    (c) the contact line friction obtained from the integral of the memory kernel (green triangles) and from the Green-Kubo like Eq.~7 (blue triangles)  as a function of the length of the contact line.
    (d) the memory time calculated as the first moment of the absolute value of the memory kernel (Eq.~\ref{eq:mem_time}) against the length of the contact line.
    }
    \label{fig:size_effect}
\end{figure}

Fig.~\ref{fig:size_effect} (a) shows the variance of contact line fluctuations $\expval{X^2}$ multiplied by the contact line length as a function of contact line length $L_y$.
The data indicates that  $\expval{X^2}$ decreases with increasing $L_y$.
This can be understood by considering a large contact line.
If we split this contact line into $n$ approximately independent segments, then the variance of contact line fluctuations decreases linearly with $n$ since we average over the segments by projecting the system into the $xz$-plane.
We find that the product of variance and contact line length depends only weakly on the contact line length.
Fig.~\ref{fig:size_effect} (b) shows the mass of the contact line divided by the length of the contact line as a function of the contact line length.
The contact line mass appears to approach a linear increase with the contact line length, but not on the length scales simulated in this work.
The error of the extracted contact line mass is significantly smaller than the symbols.
Fig.~\ref{fig:size_effect} (c) shows the contact line friction coefficient obtained from the integral of the memory kernel (green triangles) and the Green-Kubo-like Eq.~7  (blue triangles) as a function of the contact line length.
When determining the contact line friction with Eq.~7 we first obtain the position of the plateau from the averaged correlation function and then calculate an average and uncertainty estimate from all individual correlation functions.
Both approaches agree well within uncertainty. 
We find that the dependence of the contact line friction on the contact line length is weak and, within the error estimate, not statistically significant.
We do not report an uncertainty for the  contact line friction obtained from the memory kernel here, because while estimating uncertainties for correlation functions is trivial, the memory kernel extraction exponentially amplifies uncertainties on large time scales. 
This means that averaging correlation functions over ten simulations and extracting the memory kernel with the resulting correlation functions yields a result with a much smaller error than extracting a memory kernel for each simulation and estimating the uncertainty.
Fig.~\ref{fig:size_effect} (d) shows the memory time $\tau_{\mathrm{mem}}$
\begin{equation}
    \tau_{\mathrm{mem}} = \frac{1}{\int^{\infty}_0\dd t \, |K(t)|} \int^{\infty}_0\dd t \, |K(t)|t \label{eq:mem_time}
\end{equation}
computed as the first moment of the absolute value of the memory kernel as a function of the contact line length.
With differences in the low percent range, we conclude that the contact line length has a negligible effect on the memory time.
\FloatBarrier
\clearpage

\section{Additional information on the Vibrational Density of State of the Bulk Liquids}
To relate the oscillations in the VACF to molecular movements, we calculate the vibrational density of states $g(\omega)$ of the bulk liquids,
\begin{equation}
    g(\omega) = \frac{m\hat{C}(\omega)}{k_{\mathrm{B}}T} = \frac{\hat{C}(\omega)}{C(0)},
\end{equation}
such that
\begin{equation}
    g(0) = \frac{m}{k_{\mathrm{B}}T}\int^{\infty}_{-\infty}\dd t \,C(t) = \frac{2 m D}{k_{\mathrm{B}}T},
\end{equation}
where $D$ is the self-diffusion coefficient associated with the VACF (equal to zero for the contact line).

To obtain the vibrational density of states of the LJ-fluid, we prepare a system of $8000$ molecules. 
We first relax the system by integrating it for $500\,\tau_{\mathrm{LJ}}$ in the $NVT$ ensemble using a Nosé–Hoover thermostat with a thermostat constant of $0.5\,\tau_{\mathrm{LJ}}$.
To obtain an equilibrated density we integrate the system for $5000\,\tau_{\mathrm{LJ}}$ in the $NPT$ ensemble at a pressure of $1\,\epsilon\,\sigma^{-3}$ using a Nosé–Hoover style barostat with a barostat constant of $5\,\tau_{\mathrm{LJ}}$ and a Nosé–Hoover thermostat with a thermostat constant of $0.5\,\tau_{\mathrm{LJ}}$.
The average box length is $42.009\,\sigma$.
Finally we perform a production run in the $NVT$ ensemble for $500\,\tau{\mathrm{LJ}}$ using a Nosé–Hoover thermostat with a thermostat constant of $5\,\tau_{\mathrm{LJ}}$.
For the production the velocity of all beads is saved at every step.

To obtain the vibrational density of states of SPC/E water, we prepare a system of $8000$ water molecules. We first minimize the energy of the system. 
Then we thermalize the system by integrating it for $100\,\mathrm{ps}$ in the $NVT$ ensemble using a Nosé–Hoover thermostat with a thermostat constant of $0.2\,\mathrm{ps}$.
To relax the density of the system we integrate it for $2\,\mathrm{ns}$ in the $NPT$ ensemble at a pressure of $1\,$bar using a Nosé–Hoover style barostat with a barostat constant of $2\,\mathrm{ps}$ and a Nosé–Hoover thermostat with a thermostat constant $0.2\,\mathrm{ps}$.
We obtain an average box length of $62.370\,\sigma$.
Finally we perform a production run in the $NVT$ ensemble for $2\,\mathrm{ns}$ using a Nosé–Hoover thermostat with a thermostat constant of $2\,\mathrm{ps}$.
For the production run the velocity of the oxygen atoms is saved at every step.

\begin{figure}[h]
    \centering
    \includegraphics[width=\linewidth]{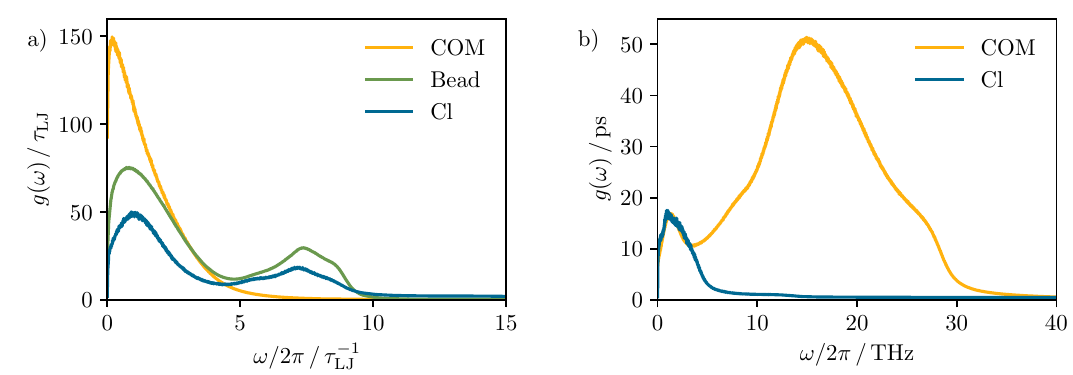}
    \caption{
    (a) the vibrational density of states of the LJ-fluid beads (green line) and the center of mass of the LJ-fluid molecules (yellow line) in the bulk fluid and the vibrational density of states of the corresponding contact line (blue line).
    The peaks in the bead and contact line spectrum overlap, indicating that oscillations in the contact line VACF are caused by bead vibrations.
    (b) the vibrational density of states of bulk SPC/E water (yellow line) and the vibrational density of states of the corresponding contact line (blue line).
    The peaks in the SPC/E and contact line spectrum overlap, indicating that oscillations in the contact line VACF are caused by molecular vibrations.
    }
    \label{fig:dos_lq}
\end{figure}

We compute the Fourier transform of the VACF via fast Fourier transforms as implemented in NumPy from the Fourier transform of the velocity.
Fig.~\ref{fig:dos_lq}(a) shows the vibrational density of states of the LJ beads (green line) and the center of mass of the molecule (yellow line).
The center of mass spectrum features a single peak at $0.2\,\tau_{\mathrm{LJ}}$. 
The Bead spectrum is richer with two prominent peaks at $0.9\,\tau_{\mathrm{LJ}}$ and $7.5\,\tau_{\mathrm{LJ}}$ overlapping with peaks in the contact line spectrum (blue line) at  $1\,\tau_{\mathrm{LJ}}$ and $7.2\,\tau_{\mathrm{LJ}}$.
Fig.~\ref{fig:dos_lq}(b) shows the vibrational density of states of SPC/E.
For the computation, we use the oxygen position as an approximation to the center of mass.
The spectrum features a prominent peak at $15\,$THz and a smaller peak at $1.5\,$THz overlapping with a peak in the contact line spectrum (blue line) at $1.2\,$THz.
We conclude that the oscillations in the contact line VACF are caused by low to intermediate-frequency molecular vibrations. 
\FloatBarrier
\clearpage

\section{Additional information on Coupling between the Contact Lines}
\begin{figure}[H] 
    \centering
    \includegraphics[width=0.5\linewidth]{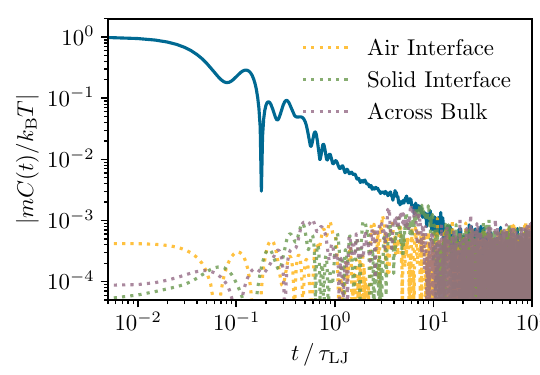}
    \caption{VACF of the contact line (solid blue line) and velocity cross-correlation along the air-liquid
    interface (dotted yellow line), along the solid-liquid interface (dotted green line), and across the bulk (dotted purple line).
    These cross-correlations being non-zero strongly indicates interactions between the contact lines via the system's hydrodynamics.}
    \label{fig:vacf_cross}
\end{figure}

Fig.~\ref{fig:vacf_cross} shows the VACF of the contact line (solid blue line) and velocity cross-correlations along the air-liquid
interface (dotted yellow line), along the solid-liquid interface (dotted green line), and across the bulk (dotted purple line) for the LJ system.
For short times, all cross-correlations tend to zero.
However, for large times of $10\mathrm{\tau}_{\mathrm{LJ}}$, the magnitude of all cross-correlations increases and approaches the magnitude of the VACF.
This indicates strongly that the contact lines interact via the hydrodynamics of the system since the distance between the contact lines is at least one order of magnitude larger than the LJ-cutoff.

\begin{figure}[H]
    \centering
    \includegraphics[width=\linewidth]{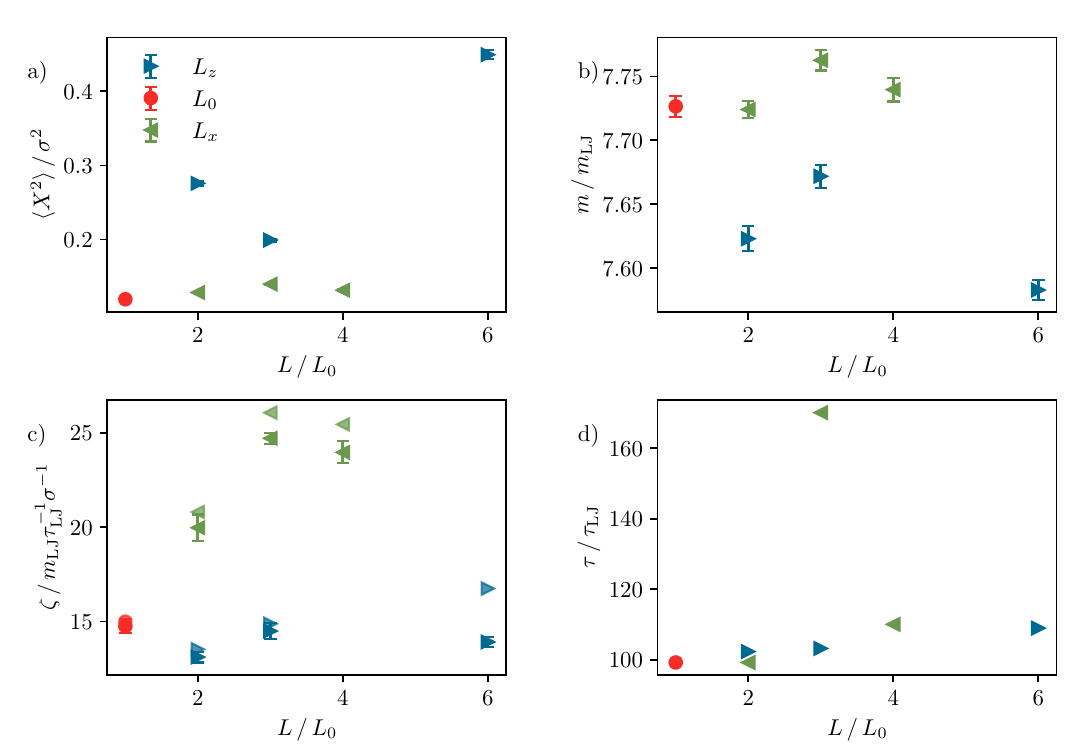}
    \caption{
    (a) the mean amplitude of the contact line fluctuations, (b) the mass of the contact line divided by the length of the contact line, (c) the contact line friction, and (d) the memory time calculated with Eq.~\ref{eq:mem_time} as a function of the dimension of the droplet along the $z$-axis (blue triangles) and along the $x$-axis (green triangles).
    The red dot indicates the system discussed in the main text.
    }
    \label{fig:cl_size_xz}
\end{figure}

Associated with this interaction are finite-size effects likely due to the restriction placed on the hydrodynamic modes by the dimensions of the fluid droplet.
Fig.~\ref{fig:cl_size_xz} (a) shows the mean amplitude of contact line fluctuations as a function of the system size along the $z$-axis (blue triangles), the $x$-axis (green triangles), and the system discussed in the main text (red dot)
Increasing the length of the fluid droplet along the $x$-axis does not result in a change of the amplitude within the estimated error, however increasing the dimensions of the fluid droplet along the $z$-axis (and thereby increasing the air-liquid interface) does. 
Note that this corresponds to a change to the systems thermodynamics and does not directly influence the dynamics.
Fig.~\ref{fig:cl_size_xz} (b) shows the mass as function of the system size. 
We find that the mass does not change within error when the dimensions of the droplet are increased along the $x$-axis and decreases by $\approx2\%$ when the dimensions of the droplet are increased along the $z$-axis.
Fig.~\ref{fig:cl_size_xz} (c) shows the contact line friction as function of the system size. 
The triangles with error bars indicate the contact line friction obtained from the Eq.~7 and the triangles with out error bars the plateau of the memory kernel.
Fig.~\ref{fig:cl_size_xz} (d) shows the memory time calculated with Eq~\ref{eq:mem_time} as a function of the system size.
Increasing the dimensions of the droplet along the $z$-axis does not result in a significant change in the contact line friction or the memory time.
In contrast we find a strong increase of the contact line friction (especially for a finite size effect) when increasing the dimension of the droplet along the $x$-axis.
While the memory time does not differ strongly from the system discussed in the main text for $L_x=200\,\sigma$ and $L_x=400\,\sigma$ the memory time for $L_x=300\,\sigma$ it is significantly larger.
\clearpage

\section{Additional information on the Calculation of Discrete Fourier Transforms}
\begin{figure}[h]
    \centering
    \includegraphics[width=0.5\linewidth]{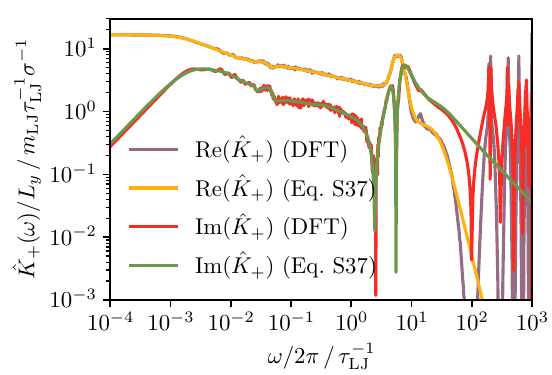}
    \caption{The Fourier transform of the memory kernel computed with DFT (purple and red lines) and the Fourier transform computed with Eq.~\ref{eq:dft_mat}  (yellow and green lines)}
    \label{fig:dft_mat}
\end{figure}

To compute the response function and frequency-dependent friction the Fourier transform of the memory kernel has to  be evaluated for arbitrary frequencies.
Fig.~\ref{fig:dft_mat} shows the discrete Fourier transform (DFT) of the memory kernel for the LJ system.
The integrals are evaluated with the Trapezoidal rule (purple line for real part and red line for imaginary part).
For high frequencies the integrals do not converge on the time grid the memory kernel is sampled on resulting in oscillations for large frequencies.
These problems can be approached with a Filon algorithm or adapted versions; however, here we utilize that the memory kernel may be represented in terms of a (diffusion) matrix $\bm{A}$
\begin{equation}
    K(t) = \bm{A}_{Py}^T e^{-\bm{A}_{yy}t} \bm{A}_{Py},
\end{equation}
where 
\begin{equation}
    \bm{A} = 
    \begin{pmatrix}
         0 & \bm{A}_{Py}^T \\
         -\bm{A}_{Py} & \bm{A}_{yy}
    \end{pmatrix}.
\end{equation}
In this matrix representation, there is an analytical solution for the (half-sided) Fourier transform of the memory kernel
\begin{align}
    \hat{K}_{+}(\omega) =& \int^{\infty}_{0}\dd t\, e^{i\omega t}\bm{A}_{Py}^T e^{-\bm{A}_{yy}t} \bm{A}_{Py} \notag\\
    =& \int^{\infty}_{0}\dd t\, \bm{I} e^{i\omega t}\bm{A}_{Py}^T \bm{I} e^{-\bm{A}_{yy}t} \bm{A}_{Py} \notag\\
    =& \bm{A}_{Py}^T \left(\int^{\infty}_{0}\dd t\, e^{-(\bm{A}_{yy}-i\omega\bm{I})t}\right) \bm{A}_{Py} \notag\\
    =& \bm{A}_{Py}^T \left(\bm{A}_{yy} -i\omega\bm{I}\right)^{-1} \bm{A}_{Py}, \label{eq:dft_mat}
\end{align}
shifting the problem to solving a least-squares problem to determine $\bm{A}$ with sufficient accuracy.
We solved this problem in a past work using a Gauss-Newton scheme.
We obtain a good fit with to the extract memory kernel with a $19\cross19$ and a $15\cross15$ matrix for the LJ and water system respectively.
Fig.~\ref{fig:dft_mat} shows the discrete Fourier transform of the memory kernel using Eq.~\ref{eq:dft_mat} (the yellow line is the real part and the green line the imaginary part).
We obtain an accurate representation of the Fourier transform for all frequencies, while avoiding numerical problems for large frequencies.
\FloatBarrier
